\documentclass[preprint,prb,nobibnotes,altaffilletter,amsmath,amssymb,amsfonts]{revtex4}

\usepackage{inputenc}
\inputencoding{ansinew}
\usepackage[T1]{fontenc}
\usepackage{graphicx}
\usepackage{palatino}
\usepackage{mathpazo}
\usepackage{bm}

\usepackage{color}

\newcommand*{\diff}[2]{\frac{\mathrm{d}#1}{\mathrm{d}#2}}
 \renewcommand*{\(}{\left(}
  \renewcommand*{\)}{\right)}
  \newcommand*{\dlog}[2]{\frac{\mathrm{d} \ln #1}{\mathrm{d} \ln #2}}
\newcommand*{\uu}{\langle u^2 \rangle}

\newcommand*{\plog}[2]{\frac{\partial \ln #1}{\partial \ln #2}}

\begin{document}

\title{Connection between slow and fast dynamics of
   molecular liquids around the glass transition}
\author{Kristine Niss$^{1,2}$, Cécile Dalle-Ferrier$^1$, Bernhard Frick$^3$, Daniela Russo$^3$, Jeppe Dyre$^3$
  and Christiane Alba-Simionesco$^{1,4}$}
\address{$^1$Laboratoire de Chimie Physique, B{\^a}timent 349, Universit{\'e} Paris-Sud, F-91405 Orsay, France\\
$^2$DNRF centre  ``Glass and Time'', IMFUFA,  Department of Sciences, Roskilde University, Postbox 260, DK-4000 Roskilde, Denmark\\
$^2$Institut Laue-Langevin, F-38042 Grenoble, France.\\
$^4$Laboratoire Léon Brillouin, UMR 12, CEA-CNRS, 91191-Gif-sur-Yvette, France.\\}
\date{\today}

\begin{abstract}
  The mean-square displacement (MSD) was measured by neutron
  scattering at various temperatures and pressures for a number of
  molecular glass-forming liquids. The MSD is invariant along the
  glass-transition line at the pressure studied, thus establishing an
  ``intrinsic'' Lindemann criterion for any given liquid.  A
  one-to-one connection between the MSD's temperature dependence and
  the liquid's fragility is found when the MSD is evaluated on a time
  scale of $\sim$4 nanoseconds, but does not hold when the MSD is
  evaluated at shorter times. The findings are discussed in terms
  of the elastic model and the role of relaxations, and the
  correlations between slow and fast dynamics are addressed. 
\end{abstract}

\maketitle{}

\section{Introduction}

A major challenge to current condensed-matter physics is to explain
properties of supercooled liquids approaching the glass transition. In
particular, there is still no consensus about what causes the dramatic
``super-Arrhenius'' increase of the liquid's viscosity. The viscosity
will for so-called fragile liquids close to the glass transition
increase with about one decade (in some cases even more) for a
temperature decrease of just 1$\%$. The most popular model is the
Adam-Gibbs entropy model \cite{ada65} which relates the slowing down
to an underlying phase transition; this model involves the notion of a
growing length scale as the liquid is cooled towards the glass
transition. The experimental search for growing length scales uses
many different observables, it has been
going on for several decades \cite{donth82,tracht98} and it is an
ongoing active field of research \cite{berthier05,dalle07,thi10}, but there is still no clear
conclusion. Another approach is advocated by the so-called elastic
models, the main ideas of which date back to a paper by Eyring and
co-workers from 1943 \cite{tob43} and Nemilov's further work in 1968
\cite{nemilov68}. These models relate the activation energy to the
viscous liquid's short-time elastic properties
\cite{dyr96,dyre06}. Thus, apparently paradoxical, properties on the
pico- or nanosecond time scale could determine the slow molecular
relaxations taking place over minutes or hours in the liquid close to
the conventional glass-transition temperature. This paper investigates
this intriguing prediction by studying several liquids, some under
varying pressure, by means of quasi-elastic neutron scattering
experiments providing the most direct measurement available of the
mean-square displacement on several short time scales.

The idea of a connection between the dynamics on time scales differing
by ten or more orders of magnitude has also been put forward in other
contexts.  If correct, it emphasizes the fact that the
glass-transition phenomenon involves an exceedingly large dynamical
range --- and that a full understanding of the glass transition must
encompass both fast and slow dynamics. Before proceeding to describe
the experiment, we give examples of the ideas and results which point
in this direction.

In 1992 it was observed by Buchenau and Zorn \cite{buchenau92} that
there is a relation between the temperature dependence of the
structural relaxation time and the temperature dependence of the
mean-square displacement (MSD) observed on the nano-pico second time
scale in selenium as determined by neutron scattering
experiments. Since then, other groups found similar results, showing
qualitatively that the larger the fragility is, the stronger is the
temperature dependence of the MSD
\cite{angell95,kan99,casalini01,ngai00,mag04,ngai04,cor05}. Recent
theoretical work relates the mean-square displacement and the
vibrational entropy and relates both quantities to the slow dynamics \cite{wyart10}.
In 1987
Hall and Wolynes \cite{hal87} theoretically discussed how the
mean-square vibrational displacement controls the relaxation time
according to the expression $\tau\propto\exp({\rm Const.}/\uu)$. Their
approach was later developed into the random first-order transition
theory (RFOT) of the glass-transition \cite{xia00,lub06} where a
variational density profile built of Gaussian vibrational
displacements around aperiodic atomic positions is optimized for
free-energy minimization. Thus the vibrational short-time displacement
is the crucial quantity for the RFOT, which was later developed into a
full-fledged theory leading to a generalization of the Adam-Gibbs prediction
for the relaxation time in terms of the configurational entropy
\cite{ada65}.

In more recent works \cite{nov04, nov05} Novikov and Sokolov
demonstrated a surprising connection between the ``fragility'' --- a
measure of how fast the liquid's viscosity (or relaxation time)
increases as temperature decreases and enters the glassy state --- and
elastic properties of the glass: The more fragile the liquid is, the
higher is the ratio between the bulk and shear moduli of the resulting
glass. However, a study of a larger set of liquids shows that the
correlation does not hold on general \cite{yanno06}. Novikov and
Sokolov discussed a possible explanation of their correlation in terms
of elastic models like the shoving model \cite{dyr96}. This does not
seem to be a correct connection because recent experimental work by
Nelson and coworkers supports the shoving model while it lends no
support to Novikov's and Sokolov's correlation \cite{tor09}. Sokolov
earlier introduced a parameter derived from the measured dynamic
structure factor, which relates the strength of the quasielastic
scattering intensity at $T_g$ normalized to the intensity of the boson
peak and the fragility\cite{sok93}. This correlation has also been
questioned \cite{yanno00} even if it holds on a very qualitative level
(intense boson peaks are seen in strong systems like oxide glasses
whereas the boson peak in van der Waals systems is less
pronounced). It also appears that this correlation could be more
related to isochoric than isobaric fragility \cite{niss06}. Along
similar lines Scopigno and coworkers suggested a correlation between
the temperature dependence of the non-ergodicity factor measured in
the glass and the fragility of the liquid \cite{scop03,scop10}. The
main point to be noted here is that the two Novikov-Sokolov
correlations as well as the Scopigno correlation provide an intriguing
connection between short- and long-time properties of the liquid
(since the liquid's short-time mechanical properties are those of the
glass corresponding to the liquid structure at the temperature in
question).

Widmer and Harrowell \cite{wid05} proposed studying the
Maxwell-Boltzmann ensemble averaged mean-squared displacement of a
particle for any given initial configuration (the
``iso-configurational ensemble''), terming this quantity the dynamic
propensity of the particle in question. This property's distribution
reflects the dynamic heterogeneity of the liquid. Thus, once again, a
connection is established between the long-time dynamic properties and
MSD on time scales much shorter than the relaxation time.

Leporini and collaborators \cite{lar08} argued from
simulations that there is a universal correlation between the
structural relaxation time and the ``rattling amplitude'' from high-
to low-viscosity states. According to this picture the glass softens
when the rattling amplitude exceeds a critical value. This implies a
``universal'' Lindemann criterion for the glass transition, i.e., that
the glass transition takes place when the MSD reaches a certain value
(see e.g. Ref. \onlinecite{sanditov03,dyre06}).

At first sight it appears very surprising that there could be any
relation between the alpha-relaxation process --- taking place on the
second or hour time scale --- and mean-square displacements taking
place on the nano/pico second time scale. It should be recalled,
however, that whereas the alpha relaxation is very slow, the barrier
transitions themselves are fast. This fact is the starting point for
the elastic models. In these models the relation between mean-square
displacement and fragility comes very natural, and the stiffness of
the material or, equivalently, the steepness of the energy minima,
determines the activation energy of the alpha process
\cite{tob43,hal87,dyre06}.

In terms of the vibrational MSD $\uu$, in the simplest version where
the instantaneous bulk and shear moduli are proportional in their
temperature variation, all elastic models imply for the activation
energy $\Delta E(T)\propto Ta^2/ \uu(T)$ at atmospheric pressure, where
$a$ is the average intermolecular distance
\cite{hal87,dyre06,ngai04,u2_ref}. This implies

\begin{eqnarray}
  \tau(T)\,=\,\tau_0\, \exp\(\frac{Ca^2}{\uu(T)}\),
  \label{eq:elas}
\end{eqnarray}
where $C$ is a constant. This result relates a larger MSD to a shorter
alpha relaxation time and it implies that the temperature dependence
of the relaxation time is governed by the temperature dependence of
the MSD. Hence the elastic model predicts that the change of the MSD
just above $T_g$ is more dramatic the more fragile the liquid is. This
agrees with the generally observed trend, although questions remain
about at which time scale $\uu$ should be considered. The
glass-transition Lindemann criterion states that the $\uu/a^2$, i.e.,
the relative vibrational amplitude of the atoms, at the glass
transition should reach a certain universal value allowing diffusion
on long time and length scales
\cite{sanditov03,buchenau04a,dyre06,lindemann}. Recall that the
Lindemann criterion is the rule that $\uu/a^2\sim$1\% when any crystal
melts. If the glass transition is also characterized by such a
universal number, there would be an appealing analogue between crystal
and glass ``melting'' -- although the latter phenomenon is known to be
cooling rate dependent.

In this paper we present a \emph{quantitative} test of
Eq.~(\ref{eq:elas}) based on MSD data in the nano- and pico-second
time scale obtained by neutron backscattering and time-of-flight
techniques on several molecular liquids, covering fragilities ranging
from $49$ to $161$. Three liquids were also studied under varying
pressure. This allows one to examine different glass transitions of
the same liquid, i.e., without changing the intermolecular
interactions.


\section{Experimental}\label{sec:exp}

The experiments were carried out on the back-scattering instruments
IN10 and IN16 at the ILL. This spectrometers use the (1~1~1)
reflection of Si-single crystals with a Bragg angle of 90$^\circ$ at
the monochromator and analyzers to reach an energy resolution of
FWHM=1 $\mu$eV (corresponding to a time scale of $\sim$4~ns). The
wavelength of the neutrons was 6.27~\AA{}. The wavevector, $Q$, range
covered was 0.2\AA$^{-1}$ to 1.9~\AA$^{-1}$. The experiments were
performed isobarically in cooling with a rate of $\sim 0.5$
K/min. Pressure was applied using a clamp pressure cell mounted on the
bottom of an insert to the cryostat. Sample transmission was 88\% for
the high-pressure measurements and $95\%$ for the measurements at
atmospheric pressure in standard aluminum cells. The liquids studied
are: glycerol, cumene, dibuthylthalate (DBP), m-toluidine, sorbitol,
triphenylphosphite (TPP) and decahydroisoquinoline (DHIQ).  The first
three were also studied at elevated pressure (300 MPa or 500 MPa); the
pressure dependence of the glass-transition temperature was obtained
from calorimetric experiments or extracted from dielectric
spectroscopy under pressure. For glycerol a different pressure cell,
dedicated for studying liquids under hydrostatic pressure, was
used. This cell is built out of Niobium resulting in a low background
and the pressure can be adjusted via a capillary from the outside of
the cryostat, but the maximum pressure is limited to 300 MPa
\cite{frick99}. The raw data correction was performed using the
standard ILL software Sqwel which converts the measured data to the
scattering law $S(Q,\omega)$ for the sample. The MSD is calculated
from the measured elastic intensities by adopting the Gaussian
approximation, $\ln (I) = A - \frac{Q^2\uu}{3}$ with $A$ being a
constant.  The measured intensity is normalized to the intensity
measured at low temperature, $T=4$~K, which means that the zero point
motion is removed. We find that the $Q^2$ dependence is obeyed in the
temperature range 0 K to 1.2 $T_g$ \cite{NissThesis,
  DalleThesis}. Even so, one should still be aware that the measured
MSD can contain local relaxations which are unrelated to the
vibrations and independent of the structural relaxation.

Supplementary experiments on DHIQ and DBP were carried out on the
back-scattering instrument IN13 at atmospheric pressure. The energy
resolution on IN13 is almost ten times wider, FWHM=8~$\mu$eV, meaning
that the MSD we access in the measurement is probed on an almost 10
times faster time scale ($\sim$0.5~ns).

\section{The Lindemann Criterion}\label{sec:linde}

Our measurements give the molecular MSD at the nano and the picosecond
time scale. How can the elastic model prediction be tested?  Comparing
data for the same liquid at differing pressures, avoids making
assumptions about the constant $C$ of Eq. \ref{eq:elas}. If it turns out that $C$ is common
to all liquids, a universal (i.e., genuine) glass-transition Lindemann
criterion is implied.

\begin{figure}[htbp]
  \centering
\includegraphics[width=15cm]{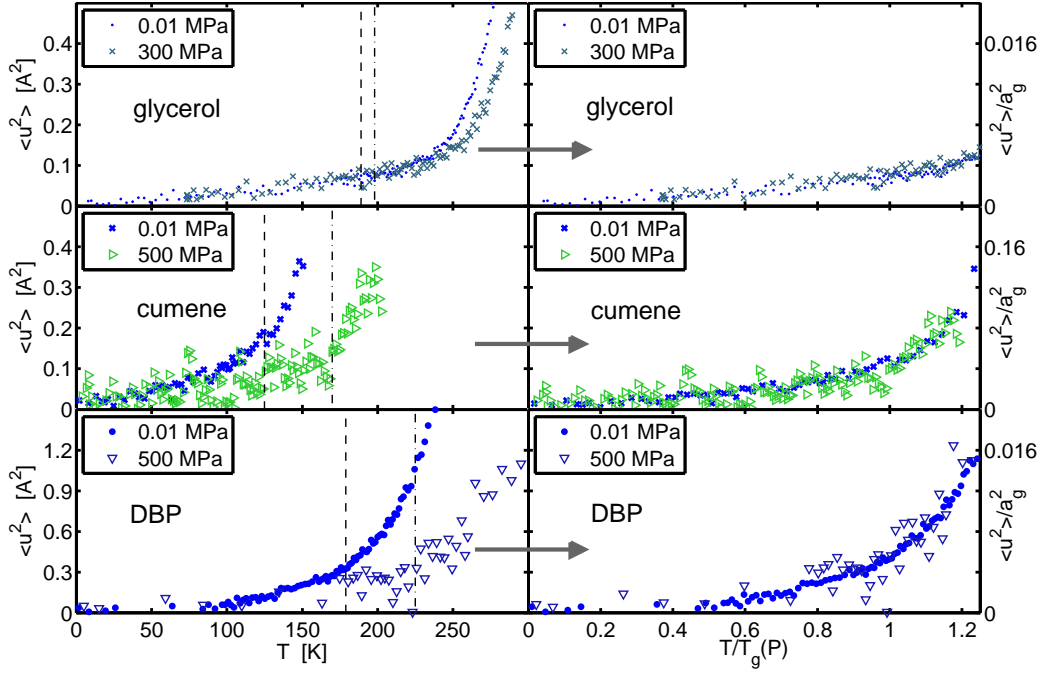}
\caption{(Color online) The MSD of glycerol, cumene and DBP at atmospheric pressure
  and at 500~MPa (300~MPa for glycerol). The left hand side of the
  figure shows $\uu$ and temperature on an absolute scale. The dashed
  lines indicate $T_g$, the dash-dotted lines $T_g$ at high
  pressure. The temperature scale in the right hand side of the
figure is scaled by the pressure dependent $T_g$ and the y-axis is scaled
  with $a^2\propto\rho^{-2/3}$ evaluated at $(T_g(P),P)$.}
\label{fig:LindeP}
\end{figure}

In Fig. \ref{fig:LindeP} we present data for three liquids of
different chemical nature, intermolecular interactions and fragility
(glycerol, cumene, and dibutylphthalate). They are studied at ambient
as well as at high pressures (300 MPa for glycerol, 500 MPa for the
two other liquids). The left part of each figure gives the mean-square
displacement as a function of temperature at the two pressures where
the dashed line marks the glass-transition temperature. The right part
gives the data scaled as implied by Eq.~(\ref{eq:elas}). Thus the
temperature is scaled by $T_g$ and the MSD by $a^2$, where we assume that
the intermolecular distance scales with $a^2\propto\rho^{-2/3}$. The
density, $\rho$ is evaluated at $(T_g(P),P)$ from known equations
of state\cite{note1}. For all three liquids there is data collapse,
showing that a Lindemann type criterion is fulfilled. 

The scaling of the temperature axis is by far the most important for
this data collapse in the pressure range studied. The estimated
increase in density is less than 10\%. This gives a decrease of $a^2$
by approximately 5\%. This difference is almost indistinguishable in
figure \ref{fig:LindeP} due to the scatter of the data. A scaling with
the pressure-dependent glass-transition temperature was earlier shown
by two of us for a polymer sample in ref. \onlinecite{frick99}. The
earlier scaling is also done at relatively low pressures where the
density does not change dramatically. It thus appears that for the
systems studied so far, the MSD is constant along the glass-transition
line in a P-T-diagram, suggesting a Lindemann criterion. To verify
whether the MSD is constant or, as suggested by the elastic model, if
the normalized MSD, $\uu(T)/a^2$, is constant studies are needed in a
larger pressure range. A constant value of $\uu(T)/a^2$ with
$a^2\propto\rho^{-2/3}$ along the glass-transition line, $(T_g(P),P)$,
is also consistent with the existence of isomorphs as it is pointed out
in Ref. \onlinecite{gnan09}. Alternatively, the change of the
intermolecular distance $a$ could be found based on a more local
measure extracted from the static structure factor, $S(Q)$. However,
measurements as a function of temperature and pressure show that the
$Q$-dependence of the peak maximum follows the same behavior
\cite{alba98}.

\begin{figure}[htbp]
  \centering
\includegraphics[width=8cm]{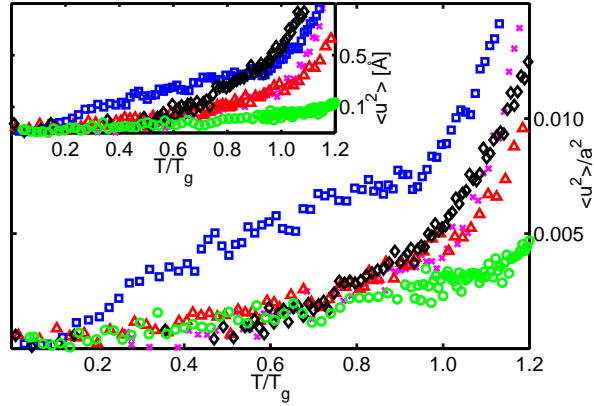}
\caption{(Color online) The temperature dependence of $\uu$ scaled to $a^2=v^{2/3}$   (see text for details) for 5 different liquids; glycerol (circles), DBP (diamonds), m-toluidine (squares), cumene (triangles), DHIQ (crosses). The temperature is scaled to $T_g$. The inset: The same data as in the main figure, here shown with the absolute value of $\uu$.} 
\label{fig:LindeA}
\end{figure}

The next step is to investigate whether the constant $C$ in
Eq.~\ref{eq:elas} is common to all liquids, as required by a universal
glass-transition Lindemann criterion \cite{dyr96,lar08}. This is
investigated in Fig. \ref{fig:LindeA} by plotting $\uu(T)/a^2$ as
function of $T/T_g$ for a selection of liquids at ambient pressure
with quite similar $T_g$'s. If the constant $C$ were universal,
$a^2/\uu$ should be the same for all liquids at $T_g$. The figure
shows that this is not the case since the number $\uu(T_g)/a^2$ varies
a factor of 3 going from glycerol to m-toluidine.  It should be noted,
though, that the temperature dependence of the MSD of m-toluidine has
a strong increase far below $T_g$. This type of behavior has earlier
been seen in other systems and is associated with the methyl-group
rotation \cite{frick94}. Such type of local motion is probably
independent of the glass-transition temperature and an irrelevant
contribution to the apparent MSD with respect to the Lindemann
criterion. Even for the four other liquids there is a factor 2 in
variation when comparing $\uu(T_g)/a^2$.  Based on these data and
assuming that for DBP, DHIQ and cumene the major contributions to the
MSD are arising from displacements which are relevant for the
structural relaxation near $T_g$, we cannot confirm the existence of a
universal Lindemann criterion as predicted by Leporini and
collaborators \cite{lar08}, at least not on the time scale we
explore. The question is also discussed in a recent publication by the
same group \cite{otto09}.

\section{Temperature dependence of the MSD above $T_g$}\label{sec:temp} 

\begin{figure}[htbp]
  \centering
\includegraphics[width=8cm]{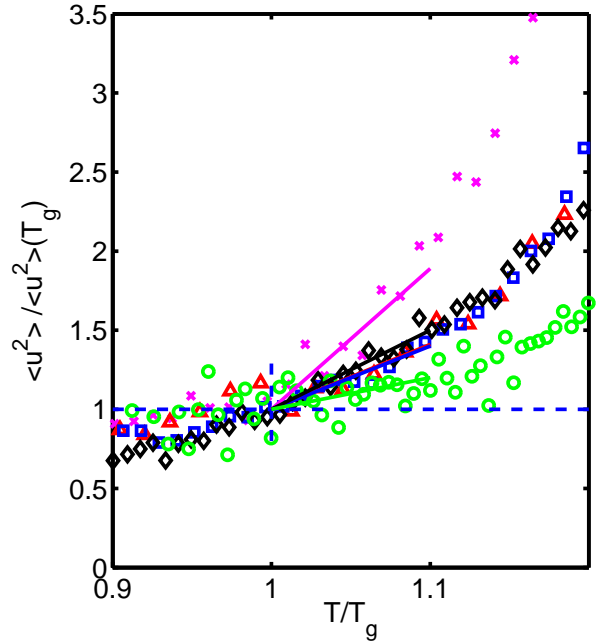}
\caption{(Color online) The measured $\uu$ scaled to $\uu_{T_g}$ for 5 different liquids. The temperature is scaled to $T_g$. glycerol (circles), DBP (diamonds), m-toluidine (squares), cumene (triangles), DHIQ   (crosses). The lines above $T_g$ illustrate the fitted slopes used in Fig. \ref{fig:predict}.} 
\label{fig:scaled}
\end{figure}

We now apply a different scaling for the MSD by
normalising to the MSD at $T_g$. Figure \ref{fig:scaled} shows $\uu(T)/\uu_{T_g}$ as a function of
$T/T_g$. The $\uu$ value of the very fragile liquid DHIQ at the
nanosecond rises most, the $\uu$ of glycerol least, dramatically; the
three remaining liquids, which all have similar intermediate
fragilities, fall in between. The systems studied hence confirm the
general trend that more fragile liquids have more temperature
dependent amplitude of the short time MSD above $T_g$ than do less
fragile \cite[]{ngai04}. The elastic models make a quantitative
prediction regarding the relation between the temperature dependence
of $\uu$ and that of the alpha relaxation time. Thus the elastic model
leading to Eq. ~(\ref{eq:elas}) is based on

\begin{equation}
  \label{eq:u2show}
\frac{\Delta E(\rho,T)}{k_B T}=\frac{C a^2}{\langle u^2 \rangle (\rho,T)} \,.
\end{equation}
Introducing the (isobaric) ``activation energy index''\cite{dyre04},
$I_P=-\left.\dlog{\Delta E(T,\rho)}{T}\right|_P$, it follows that the
elastic models predict (where the weak temperature dependence of $a$
at constant pressure is ignored)

\begin{eqnarray}\label{prediction}
  \label{}
  I_P&=&-1+\left.\diff{\ln\uu}{\ln T}\right|_{P}\,.
\end{eqnarray}
Using the general relation\cite{dyre06} between the conventional fragility index and $I_P$  it follows that the model predicts a proportionality between Angell's (isobaric) fragility $m_P$ and the relative change of $\uu$ with relative change in temperature:

\begin{eqnarray}
m_P&=& \log_{10}\(\frac{\tau_g}{\tau_0}\)\(1+I_P\)=
\log_{10}\(\frac{\tau_g}{\tau_0}\)\left.\diff{\ln \uu}{\ln
    T}\right|_{P},
\label{eq:pre}
\end{eqnarray}
where $\tau_0=10^{-14}$~s is the microscopic time and $\tau_g$=100~s
is the relaxation time at the glass-transition temperature (where
fragility is evaluated). Hence the elastic model predicts a
correspondence between the slope seen in Fig. \ref{fig:LindeA} at
$T_g$ and the fragility found from the temperature dependence of the
alpha relaxation time.

\begin{figure}
\includegraphics[width=10cm]{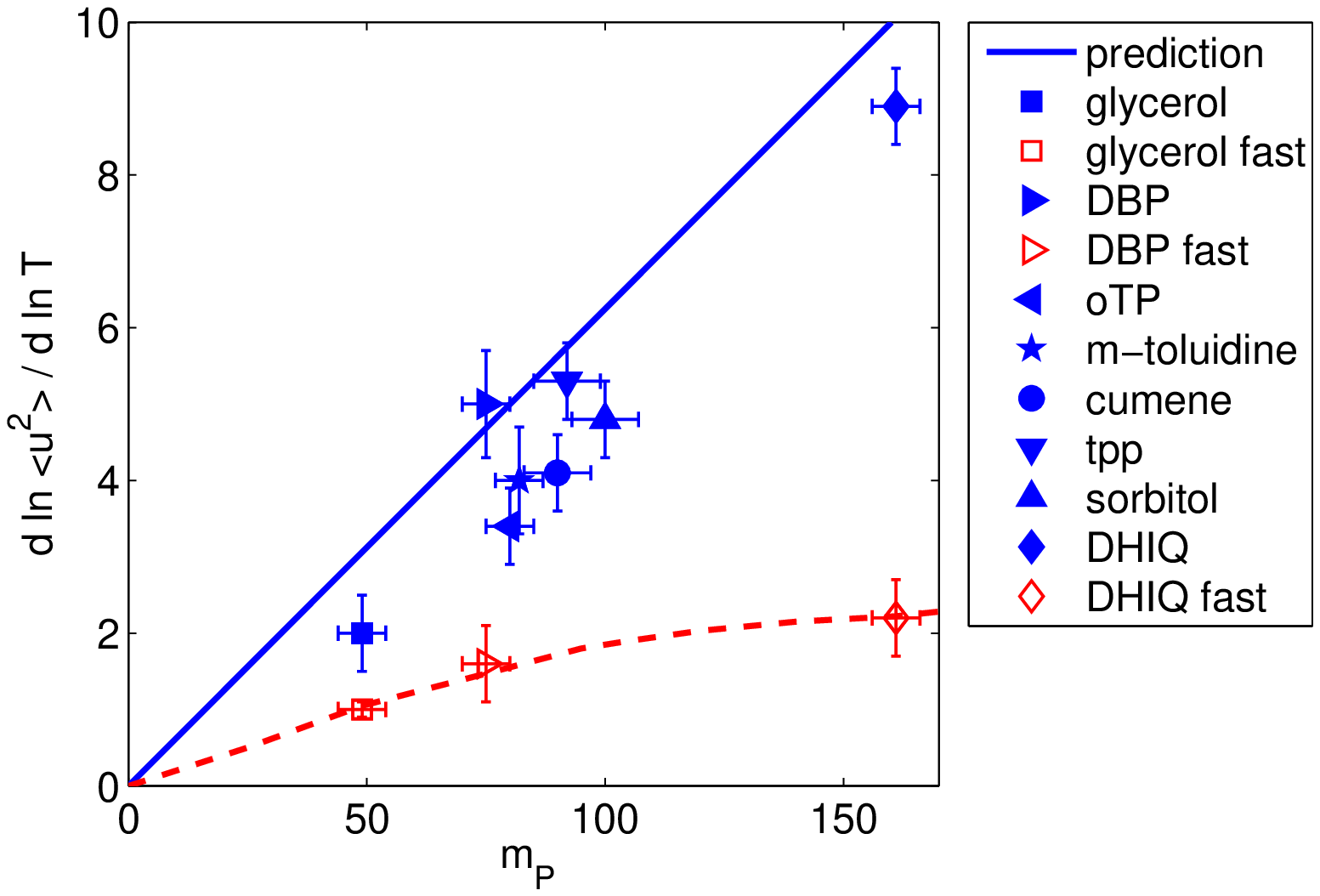}
\caption{(Color online) The value of $\left.\plog{\uu}{T}\right|_{P}$ as a function
  of the isobaric fragility. Full symbols (blue) are data that refer
  to the nanosecond time scale ($\sim$4~ns), open symbols (red) are
  data obtained on IN13 on a ten times faster time scale
  ($\sim$0.4~ns). The blue full line shows the expression
  Eq.~(\ref{eq:pre}) which follows from assuming $\Delta E(T) \propto
  T/\uu $ as is predicted from the elastic model (using $\tau_g=100$~s
  and $\tau_0=10^{-14}$~s.). Values and references are given in table
  \ref{tab}. The points calculated at $\sim$4~ns fall close to the
  predicted line, while the points calculated at $\sim$0.4~ns where
  the $\uu$ is less influenced by relaxations fall much below the line
  (the red dashed curve is a guide to the eye). When points lie below
  the blue full line, the temperature dependence of $\uu$ underestimates
  the temperature dependence of the activation energy. It should be
  noted that both $x$ and $y$ variables have considerable uncertainty
  as they are arrived at as numerical derivatives of data. The error
  bars on the literature data are our estimate based on experience
  and on the difference in reported values from different sources. }
\label{fig:predict}
\end{figure}

Figure \ref{fig:predict} tests this relation using fragilities and
$T_g$'s taken from literature (see table \ref{tab} for values and
references).  The value of $ \dlog{\uu}{T}|_{P}(T=T_g)$ is in all cases
calculated in the temperature range from $T_g$ to $\sim$1.1~$T_g$,
corresponding to the range where the fragility is determined. The data
taken on the nanosecond time scale all lie close to the line. This
result is rather convincing, especially because Eq.~(\ref{eq:pre}) not
only predicts that there is a proportionality between $m_P$ and $
\dlog{\uu}{T}|_{P}(T=T_g)$, but the value of the proportionality
constant as well.

\begin{table}[h!!!]\label{table}
\begin{tabular}{|c|c|c|c|c|c|}
  \hline
  Compound & $m_{P}$ & Refs. & $\dlog{\uu}{T}$ &$\dlog{\uu}{T}$ fast& Refs.  \\
\hline
  glycerol & 40, 53, 54 &
  \cite{alba04}$^,$\cite{birge86}$^,$\cite{paluch02} &  2  & 1 & this
  work \cite{wuttke95}\\
\hline
  DBP & 75 & \cite{niss07} & 5 & 1.6  &this work\\
\hline
  \emph{o}-terphenyl & 82, 81, 76, 84 &
  \cite{alba04}$^,$\cite{dixon88}$^,$\cite{huang01}$^,$\cite{paluch01} &
  3.4$^*$ & &\cite{casalini01}\\
  \hline
  \emph{m}-toluidine & 79,84 & \cite{alba99}$^,$\cite{mandanici05} & 4 & &this work\\
\hline
cumene & 90$^*$ & \cite{barlow66} & 4.1 & & this work\\
\hline
TPP &92$^*$ &\cite{olsen01} & 6 & &this work \\
\hline
sorbitol & 100   &\cite{olsen98} & 4.8 & &this work \\
\hline
DHIQ & 158,163$^*$ & \cite{richert03}$^,$\cite{casalini06} & 6 & 2.2 &this work\\
\hline
\end{tabular}\label{tab}
\caption{Values and references for the points shown in figure 4. The
  asterisk indicates that the value has been calculated from data in
  the paper.   
}
\end{table}

\section{The Role of Relaxations and Anharmonicity}\label{sec:anharm}

Although Fig. \ref{fig:predict} shows an overall agreement with the
elastic model prediction, when using the MSD on the nano second time
scale, a number of issues remain to be considered. Not only, of
course, is a more extensive study of different liquids needed, there
are also other more fundamental issues. One problem is that in elastic
models it is usually assumed that the measured $\uu$ is purely
vibrational, i.e., that no relaxational motion contributes to $\uu$
around $T_g$. It is not likely that this assumption is generally
correct, however. Thus we know from time-of-flight spectra that DHIQ
has a strong quasi-elastic scattering already at $T_g$
\cite{NissThesis}. Time-of-flight measurements have a broader
resolution function, and consequently shorter time scale, so this
quasi-elastic scattering corresponds to relaxation at even shorter
times than the MSD probed by backscattering. The alpha relaxation
itself also enters the experimental window at some time, possibly
already when $\tau_\alpha \sim 1 \mu$s if the relaxation function is
very stretched. This happens intrinsically faster for fragile liquids
than for strong liquids (for which, also, the relaxation functions are
generally less stretched).

When considering relaxation it also appears that the finding of
Fig. \ref{fig:predict} is consistent with another phenomenological
feature observed in the dynamic structure factor as measured from
inelastic scattering. Namely the observation that the relative
strength of the boson peak compared to the fast relaxation, measured
at $T_g$, is related to the isobaric fragility of the glass former: a
parameter, $R$, is defined as the quasielastic intensity divided by
the boson peak intensity, and proposed to increase with increasing
isobaric fragility\cite{sok93, nov05}. This model-independent
assertion made by comparing the behavior of different glass formers is
controversial\cite{yanno00}, but appears more robust than many other
correlations\cite{NissThesis,niss06,niss08}. Our findings are clearly
consistent with this observation, showing the importance of fast
processes on a time scale of a few nanoseconds even close to $T_g$,
where $R$ is determined for the most fragile liquids. This observation
at the same time suggests that the boson peak intensity itself is not
the relevant quantity for the correlation, but that it probably is the
larger intensity of fast relaxation in fragile liquids that yields the
correlation.

\begin{figure}
\includegraphics[width=8cm]{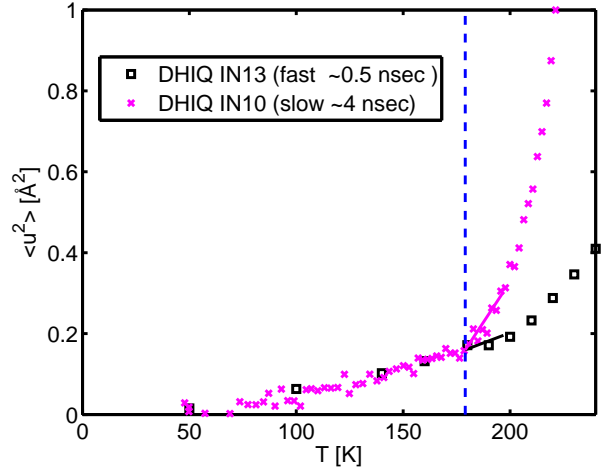}
\caption{(Color online) The MSD of DHIQ measured by IN13 (on a time scale of   $\sim$0.4~nsec) and IN10 (on a time scale of $\sim$4~nsec).  The dashed lines indicate $T_g$. The lines above $T_g$ illustrate the fitted slopes used in Fig. \ref{fig:predict}. The temperature dependence of the MSD is time scale independent below $T_g$ while it becomes strongly time scale dependent above $T_g$. This indicates that $\uu$ below $T_g$ is dominated by vibrations while relaxations play a role on the nanosecond time scale in the liquid above $T_g$ (even close to $T_g$ where the alpha relaxation time is still on the order of seconds).} 
\label{fig:timescales}
\end{figure}

To investigate the role of relaxations further we have performed
supplementary measurements of the MSD of DHIQ and DBP ( for glycerol
we used literature data\cite{wuttke95} referring to the same
timescale) using IN13 which has a broader than IN10 resolution and
therefor accesses $\uu$ on a time scale which is approximately 10
times shorter. In Fig. \ref{fig:timescales} we compare the mean-square
displacement of DHIQ found on the two different instruments, two
distinct timescales. The measured $\uu$ follow each other below $T_g$,
which strongly indicates that we probe genuine vibrations in this
regime and therefore that the finding of the Lindemann criterion is
related to the vibrations, as predicted by the elastic models. Above
$T_g$, on the other hand, we see a separation of the two curves. It is
evident that the temperature dependence of the MSD on the nanosecond
timescale probed by IN10 is much more pronounced than the temperature
dependence on the shorter time scale probed by IN13. This dependence
on the timescale indicates that we are not probing the purely
vibrational MSD on the nanosecond time scale, but rather a combination
of vibration and fast relaxations.

To illuminate how the relaxations affect the value of $d\ln\uu/d\ln T$
we adopt a simple ``jump-diffusion'' type modelling\cite{singwi60}: If
on the time scale set by the experiment some molecules vibrate whereas
others jump once or more, the MSD separates into two contributions:
$\uu=\uu_{\rm vib}+ \uu_{\rm jump}$.  For the log-log derivatives one
finds $d\ln\uu/d\ln T= A\, d\ln\uu_{\rm vib}/d\ln T + B\, d\ln\uu_{\rm
  jump}/d\ln T$ where $A=\uu_{\rm vib}/\uu$ and $B=\uu_{\rm jump}/\uu$
give the relative weights of the two contributions ($A+B=1$). The jump
contribution is most likely strongly temperature dependent. Thus any
correction for this in order to get the pure elastic contribution to
$d\ln\uu/d\ln T$ pushes the points in Fig. \ref{fig:predict}
downwards, i.e., further away from the line. This is exactly what we
see for the $d\ln\uu/d\ln T$ obtained from the IN13 data on DHIQ,
which is also shown in Fig. \ref{fig:predict}. Similarly we see
that the $d\ln\uu/d\ln T$ calculated from glycerol data taken on IN13
reported by Wuttke\cite{wuttke95} lie below the line. However, the
difference between the two timescales is much less dominant for
glycerol than for the very fragile DHIQ, indicating that the
relaxation are more dominant in the latter.

Based on the above considerations, we conclude that using the
vibrational part of the MSD as deviced by the elastic models
(Eq.~(\ref{eq:elas})) underestimates the temperature dependence of the
activation energy. The reason for this could be that Eq.~\ref{eq:elas}
is based on a simplified reasoning that basically ignores
anharmonicities \cite{dyre06}. There are two nontrivial assumptions
going into this reasoning: a) The harmonic approximation, according to
which the curvature at the minimum is inversely proportional to $\uu$;
b) The energy barrier being proportional to the curvature at the
minimum, as it would be if the potential were parabolic and scale
accordingly. The first approximation applies to a good approximation
at sufficiently low temperatures and may well apply to highly viscous
liquids because these have fairly large energy barriers. The second
approximation implies that if the barrier goes to zero, so does the
curvature. However, this is not necessarily the case. Thus the simple
elastic model assumption that the barrier scales with curvature may
break down. In summary, for a given temperature dependence of $\uu$ a
more realistic model might well predict larger $d\ln\Delta E/d\ln T$
than predicted by the elastic models. This corresponds to lowering the
slope of the theoretical line of Fig. \ref{fig:predict}. This might
explain why the $d\ln\uu/d\ln T$ measured at short times where we
expect vibrations to be dominant lie on the lower side of the line.

\section{Discussion and Conclusion}\label{sec:concl}

The liquids studied show no universal glass-transition Lindemann
criterion when we compare the MSD on the nanosecond time scale. Three
of the liquids were studied at two different pressures. They obey a
pressure-dependent Lindemann criterion, as predicted by the elastic
models. Thus the use of pressure reveals a connection which is
probably masked by details in the molecular interactions and
geometries when comparing different liquids; this suggests the
existence of an \emph{intrinsic} Lindemann criterion for each
substance at their pressure dependent glass transition. More extensive
studies including several pressures along the glass-transition line of
each liquid are needed in order to establish the range in which this
result holds. A failure of a universal Lindemann criterion might also
be due to the fact that relaxations contribute to the MSD as was
discussed in the previous section. Similar to the elastic model the
Lindemann criterion is based on a vibrational picture.

Above $T_g$ it appears that the elastic-model prediction
underestimates the activation energy temperature dependence. We
suggest that these deviations are caused by anharmonic effects. In
this context it should be noted that despite the simple ``harmonic''
appearance of the elastic models, anharmonicities must play a role,
even in the simplest elastic models. Thus the effective,
temperature-dependent elastic constant (or curvature at energy minima)
reflects anharmonicity because in truly harmonic potentials the
elastic constants are temperature independent.

While the vibrational part of the mean-square displacement does not
follow the prediction of the elastic models, we find that the total
MSD measured, $\uu (T)$, at the nanosecond time
scale (vibrations and relaxations) approximately follows a
proportionality of the type $\Delta E(T)\propto T/\uu $, where $\Delta
E(T)$ is the activation energy governing the alpha relaxation. This
one-to-one finding is based on measurements of MSD on the nano second
timescale by neutron scattering as function of temperature of
molecular liquids covering a significant range of fragilities. The
proportionality $\Delta E(T)\propto T/\uu $ shows that there is a
connection between the fast and the slow dynamics close to the glass
transition. It is not clear how causal the relation is, whether the
increase in MSD leads to higher mobility and
consequently a speed up of the alpha relaxation, or the increased
MSD is a due to a precursor of alpha relaxation
itself, for example as a high
frequency von Schweidler regime.

To summarize we find (i) an \emph{intrinsic} Lindemann criterion for
each liquid as predicted by the elastic models by studying the same
liquids at different pressures; (ii) the temperature dependence of the
MSD on the nanosecond time scale links to the liquid fragility as
predicted by the elastic models. These observations are rationalized
by introducing an anharmonicity in the elastic models, and they are
fully consistent with other experimental features and correlations
found in the literature. The findings in this work underline that a
full understanding of the viscous slowing down must involve both the
fast and the slow dynamics, and suggest that elastic models offer a
starting point for understanding the connection between these
different time scales.


\begin{thebibliography}{29}
\bibitem{note1} The densities are obtained by extrapolation of literature
  data \cite{bridgman32,bridgman49,barlow66,cook94} using the well-known Tait equation and calculated expansion
  coefficients \cite{NissThesis,DalleThesis}.
\end{thebibliography}

\end{document}